\documentclass[structabstract]{aa}  
\usepackage{graphicx}
\usepackage{natbib}
\usepackage{lscape}
\usepackage{txfonts}
\def\macc   {$\dot{M}_{\rm acc}$}
\def\lacc   {$L_{\rm acc}$}
\def\lstar   {$L_*$}

\def\msun {$M_{\odot}$}
\def\lsun {$L_{\odot}$}

\begin{document}
   \title{Accurate determination of accretion and photospheric parameters in Young Stellar Objects:
the case of two candidate old disks in the Orion Nebula Cluster \thanks{This work is based on observations collected at the European Southern Observatory, Chile (Pr.Id. 288.C-5038, 089.C-0840, 090.C-0050). }}

   \author{C.~F.~Manara\inst{1},  G.~Beccari\inst{1}, N.~Da Rio\inst{2},  G.~De Marchi\inst{2}, A.~Natta\inst{3,4}, L.~Ricci\inst{5}, M.~Robberto\inst{6}, L.~Testi\inst{1,3,7}}

   \institute{European Southern Observatory, Karl Schwarzschild Str. 2, 85748 Garching, Germany\\
              \email{cmanara@eso.org}
	\and
		European Space Agency, Keplerlaan 1, 2200 AG Noordwijk, The Netherlands
         \and
             INAF - Osservatorio Astrofisico di Arcetri, Largo E.Fermi 5, I-50125 Firenze, Italy
	\and
	School of Cosmic Physics, Dublin Institute for Advanced Studies, 31 Fitzwilliam Place, Dublin 2, Ireland
	\and
		California Institute of Technology, 1200 East California Boulervard, 91125 Pasadena, CA, USA
	\and
	Space Telescope Science Institute, 3700 San Martin Dr., Baltimore MD, 21218, USA
	\and
	Excellence Cluster Universe, Boltzmannstr. 2, 85748 Garching, Germany
		}

   \date{Received May, 9th 2013; accepted July, 26th 2013}

 
  \abstract
   {Current planet formation models are largely based on the observational constraint that protoplanetary disks have lifetime $\sim$3 Myr. Recent studies, however, report the existence of pre-Main-Sequence stars with signatures of accretion (strictly connected with the presence of circumstellar disks) and photometrically determined ages of 30 Myr, or more. 
}
   {Here we present a spectroscopic study of two major age outliers in the Orion Nebula Cluster. We use broad band, intermediate resolution VLT/X-Shooter spectra combined with an accurate method to determine the stellar parameters and the related age of the targets to confirm their peculiar age estimates and the presence of ongoing accretion.}
   {The analysis is based on a multi-component fitting technique, which derives simultaneously spectral type, extinction, and accretion properties of the objects. With this method we confirm and quantify the ongoing accretion. From the photospheric parameters of the stars we derive their position on the H-R Diagram, and the age given by evolutionary models. Together with other age indicators like the lithium equivalent width we estimate with high accuracy the age of the objects.}
   {Our study shows that the two objects analyzed are not older than the typical population of the Orion Nebula Cluster. Our results show that, while photometric determination of the photospheric parameters are an accurate method to estimate the parameters of the bulk of young stellar populations, those of individual objects with high accretion rates and extinction may be affected by large uncertainties. 
Broad band spectroscopic determinations should thus be used to confirm the nature of individual objects. 
}
   {The analysis carried out in this paper shows that this method allows us to obtain an accurate determination of the photospheric parameters of accreting young stellar objects in any nearby star-forming region. We suggest that our detailed, broad-band spectroscopy method should be used to derive accurate properties of candidate old and accreting young stellar objects in star forming regions. We also discuss how a similarly accurate determination of stellar properties can be obtained through a combination of photometric and spectroscopic data. }


\authorrunning{Manara et al.}
\titlerunning{Accurate determination of accretion and photospheric parameters in Young Stellar Objects}
   \maketitle
%

\section{Introduction}
The formation of planetary systems is strongly connected to the presence, structure, and evolution of protoplanetary disks in which they are born. In particular, the timescale of disk survival sets an upper limit on the timescale of planet formation, becoming a stringent constraint for planet formation theories \citep{Haisch01, Wolf12}. As the observed timescale for the evolution of the inner disk around pre-Main Sequence (PMS) stars is of the order of a few Myr \citep[e.g.][]{Williams11}, all models proposed to explain the gas giant planet formation (core accretion, gravitational instability) are generally constrained to agree with a disk lifetime of much less than 10 Myr. 

Observations of nearby young clusters (age $\lesssim$ 3 Myr) show the presence of few outliers with derived isochronal ages of more than 10 Myr. For example, this is found in the Orion Nebula Cluster (ONC) \citep{DaRio10,DaRio12} and in Taurus \citep{White05}. The age of the objects is derived from their position on the H-R Diagram (HRD), and thus subject to several uncertainties, either observational (e.g. spectral type, extinction), intrinsic to the targets (e.g. strong variability, edge-on disk presence, \citealt{Huelamo10}), or related to the assumed evolutionary models \citep[e.g.][]{DaRio10b, Baraffe10, Barentsen11}. 
On the other hand, observations of far (d$>$1 kpc) and extragalactic more massive clusters revealed the presence of a large population of accreting objects older (ages$>$30 Myr) than the typical ages assumed for the cluster ($\sim$ 3-5 Myr). Example of these findings are the studies of NGC 3603 \citep{Beccari10}, NGC 6823 \citep{Riaz12}, and 30 Dor and other regions of the Magellanic Clouds \citep{DeMarchi10, DeMarchi11, Spezzi12}. Observations in these clusters are dominated by solar- and intermediate-mass stars, for which the age determination is subject to the same uncertainties as for nearby clusters, but also to other uncertainties related, e.g., to the fact that the age of these more massive targets is more sensitive to assumptions on the birthline \citep{Hartmann03}.   

These findings challenge the present understanding of protoplanetary disk evolution, and can imply a entirely new scenario for the planet formation mechanism. On one side, the existence of one or few older disks in young regions does not change the aforementioned disk evolution timescales, but represents a great benchmark of a possible new class of objects where planet formation is happening on longer timescales. On the other side, the presence of a large population of older objects, representing a significant fraction of the total cluster members, could imply a totally different disk evolution scenario in those environments.

We thus want to verify the existence and the nature of older and still active protoplanetary disks in nearby regions. For this reason we have developed a spectroscopic method which allows to consistently and accurately determine the stellar and accretion properties of the targeted objects. Here we present a pilot study carried out with the ESO/VLT X-Shooter spectrograph targeting two major age outliers with strong accretion activity in the ONC. 
This is an ideal region for this study, given its young mean age ($\sim$2.2 Myr, \citealt{Reggiani11}), its vicinity ($d$ = 414 pc, \citealt{Menten07}), the large number of objects (more than 2000, \citealt{DaRio10}), and the large number of previous studies \citep[e.g.][]{Hillenbrand97,DaRio10,DaRio12, Megeath12, Robberto13}. Our objectives are therefore a) to verify the previously derived isochronal ages of these two objects by using different and more accurate age indicators, and b) to assess the presence of an active disk with ongoing accretion.

The structure of the paper is the following. In Sect.~\ref{sect::obs} we report the targets selection criteria, the observation strategy, and the data reduction procedure. In Sect.~\ref{sect::method} we describe the procedure adopted to derive the stellar parameters of the objects, and in Sect.~\ref{sect::results} we report our results. In Sect.~\ref{sect::discussion} we discuss the implications of our findings. Finally, in Sect.~\ref{sect::conclusion} we summarize our conclusions.

\section{Sample, observations, and data reduction}
\label{sect::obs}

\subsection{Targets selection and description}

We have selected the two older PMS candidates from the sample of the HST Treasury Program on the ONC \citep{Robberto13}. Our selection criteria were: clear indications of ongoing accretion and of the presence of a protoplanetary disk, an estimated isochronal age much larger than the mean age of the cluster, i.e. $\gtrsim$ 30 Myr, location well outside the bright central region of the nebula in order to avoid intense background contamination, i.e. at a distance from $\theta^1$ Orionis C larger than 5\arcmin ($\sim$0.7 pc), and with low foreground extinction ($A_V\lesssim$ 2 mag). In order to find the best candidates, we have combined together HST broad-band data \citep{Robberto13} with narrow-band photometric and spectroscopic data \citep{Hillenbrand97,Stassun99,DaRio10, DaRio12}, and with mid-infrared photometric data \citep{Megeath12}. According to \citet{DaRio10,DaRio12}, the total number of PMS star candidates in the ONC field with derived isochronal age $\gtrsim$ 10 Myr, derived using various evolutionary models \citep{DAntona,Siess00, Palla, Baraffe98}, and including both accreting and non-accreting targets is $\sim$165, which corresponds to $\sim$10\% of the total population. Among these, $\sim$ 90 objects ($\sim$5\%) have ages $\gtrsim$30 Myr. The presence of ongoing accretion has been estimated through the H$\alpha$ line equivalent width (EW$_{\rm H\alpha}$) reported in \citet{DaRio10}, using as a threshold value EW$_{\rm H\alpha} > 20 $ \AA, which implies not negligible mass accretion rates ($\dot{M}_{\rm acc} \gtrsim 10^{-9} M_\odot$/yr). This is a rather high threshold that leads to select only stronger accretors. Indeed, following e.g. \citet[][]{WB03} and \citet{Manara13}, objects with EW$_{\rm H\alpha} < 20$ \AA \ can still be accretors but with a lower accretion rate. The available {\it Spitzer} photometry \citep{Megeath12} has been used to confirm the presence of optically thick circumstellar disk, already suggested by the strong H$\alpha$ emission, looking at the spectral energy distribution (SED) of the targets to see clear excess with respect to the photospheric emission in the mid-infrared wavelength range. Among the objects with isochronal age $\gtrsim$10 Myr there are 10 objects ($<$1\%) showing very strong H$\alpha$ excess, i.e. EW$_{\rm H\alpha} > 20 $ \AA. Very striking, two PMS star candidates with isochronal age $\gtrsim$30 Myr have clear indications of H$\alpha$ excess and of presence of the disk from IR photometry. These two extreme cases of older-PMS star candidates are selected for this work. 

\begin{figure}
\centering
\includegraphics[width=0.5\textwidth]{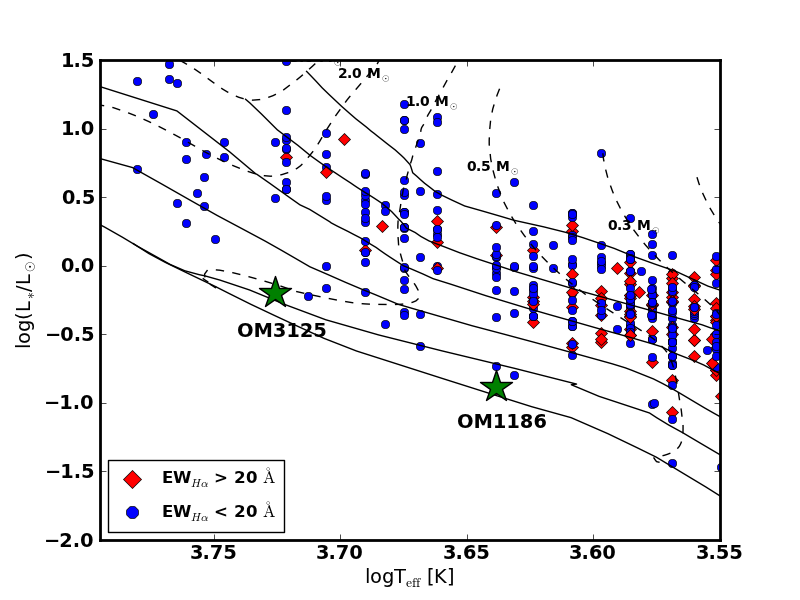}
\caption{H-R Diagram of the ONC from \citet{DaRio12}, with green stars showing the positions of the two targets of this study. The overplotted evolutionary tracks are from \citet{DAntona}. We plot (from top to bottom) the 0.3, 1, 3, 10, 30, and 100 Myr isochrones.  } 
        \label{fig::HRD}
\end{figure}

The two targets selected are OM1186 and OM3125, and we report in Table~\ref{tab::lit} their principal parameters available from \citet[][and reference therein]{DaRio10}. Both targets are located on the HRD almost on the main sequence, as we show in Fig.~\ref{fig::HRD}, where the position of these sources is reported using green stars, and we also plot the position of other PMS star candidates in the ONC field, shown with blue circles if their EW$_{\rm H\alpha} < 20 $ \AA, and with red diamonds if EW$_{\rm H\alpha} \ge 20 $ \AA. Their position on the HRD clearly indicates that they have an isochronal age much larger than the bulk of the ONC population, with ages between $\sim$ 60 Myr and  $\sim$ 90 Myr for OM1186, according to different evolutionary tracks, and between $\sim$ 25 Myr and $\sim$ 70 Myr for OM3125. The spectral type (SpT) of the two targets has been determined in \citet{Hillenbrand97} to be K5 for OM1186 and G8-K0 for OM3125. 

\begin{table*}
\centering
\caption{\label{tab::lit}Parameters available in the literature for the objects analyzed in this work}
\begin{tabular}{llcccccccc}
\hline\hline
Name & Other  & RA & DEC & SpT   & T$_{\rm eff}$&A$_V$ & L$_*$ & M$_*$& age \\
\hbox{} &names & (J2000) & (J2000)& \hbox{}  &  [K]  & [mag] & [L$_\odot$] & [M$_\odot$] &[Myr] \\
\hline
OM1186 & MZ Ori & 05:35:20.982 & -05:31:21.55&         K5 & 4350&2.40 & 0.13 & 0.66 & 64 \\
OM3125 & AG Ori & 05:35:21.687 &-05:34:46.90 &         G8 & 5320&1.47 & 0.63 & 0.95 & 25 \\
\hline
\end{tabular}
\tablefoot{Data taken from \citet[][and references therein]{DaRio10}. }
\end{table*}

\subsection{Observations and data reduction}
Observations with the ESO/VLT X-Shooter spectrograph have been carried out in service mode between February and March 2012  (ESO/DDT program 288.C-5038, PI Manara). This instrument covers simultaneously the wavelength range between $\sim$300 nm and $\sim$ 2500 nm, dividing the spectrum in three arms, namely the UVB arm in the region $\lambda\lambda\sim$ 300-550 nm, the VIS arm between $\lambda\lambda\sim$ 550-1050 nm and NIR from $\lambda\sim$ 1050 nm to $\lambda\sim$ 2500 nm \citep{Vernet11}. The targets have been observed in slit-nodding mode, in order to have a good sky subtraction also in the NIR arm. For both targets we used the same slit widths (0.5\arcsec \ in the UVB arm, 0.4\arcsec \ in the other two arms) and the same exposure times (300s x 4 in all three arms), to ensure the highest possible resolution of the observations (R=9100, 17400, 10500 in the UVB, VIS and NIR arms) and enough S/N in the UVB arm. The readout mode used was in both cases ``100,1x1,hg". The seeing conditions of the observatory during the observations were 1\arcsec \ for OM1186 and 0.95\arcsec \ for OM3125.

Data reduction has been carried out using the version 1.3.7 of the X-Shooter pipeline \citep{Modigliani}, run through the {\it EsoRex} tool. The spectra were reduced independently for the three spectrograph arms. The pipeline takes into account, together with the standard reduction steps (i.e. bias or dark subtraction, flat fielding, spectrum extraction, wavelength calibration, and sky subtraction) also the flexure compensation and the instrumental profile. Particular care has been paid to the flux calibration and telluric removal of the spectra. Flux calibration has been carried out within the pipeline and then compared with the available photometry \citep{Robberto13} to correct for possible slit losses, checking also the conjunctions between the three arms. The overall final agreement is very good. Telluric removal has been carried out using the standard telluric spectra that have been provided as part of the X-Shooter calibration plan on each night of observations. The correction has been accomplished using the IRAF\footnote{IRAF is distributed by National Optical Astronomy Observatories, which is operated by the Association of Universities for Research in Astronomy, Inc., under cooperative agreement with the National Science Foundation.} task {\it telluric}, carefully normalizing the telluric standard spectra and removing the photospheric absorption lines with a multigaussian fitting procedure.

\section{Method}
\label{sect::method}

The determination of SpT and stellar properties in accreting young stellar objects (YSOs) is not trivial for a variety of reasons. Firstly, YSOs are usually still embedded in their parental molecular cloud, which 
originates differential reddening effects in the region. This, together with the presence of a circumstellar disk surrounding the star, can modify the actual value of the extinction ($A_V$) on the central YSO from one object to another. Secondly, YSOs may still be accreting material from the protoplanetary disk on the central star. This process affects the observed spectrum of a YSO producing excess continuum emission in the blue part of the spectrum, veiling of the photospheric absorption features at all optical wavelengths, and adding several emission lines \citep[e.g.][]{Hartmann98}. The two processes modify the observed spectrum in opposite ways: extinction towards the central object suppresses the blue part of the spectrum, making the central object appear redder, thus colder, while accretion produces an excess continuum emission which is stronger in the blue part of the spectrum, making the observed central object look hotter.

For these reasons, SpT, $A_V$ and accretion properties should be considered together in the analysis of these YSOs. Here we present the minimum $\chi^2_{\rm like}$ method we use to determine SpT, $A_V$, and the accretion luminosity (\lacc) simultaneously. With this procedure we are able to estimate $L_*$, which is used to derive $M_*$ and the age of the target using different evolutionary models, and the mass accretion rate (\macc). Finally, we determine the surface gravity (log$g$) for the input object through comparison with synthetic spectra.

\subsection{Multi-component fit: set of parameters}
\label{subsec::fit}

To fit the optical spectrum of our objects we consider three components: a set of photospheric templates which will be used to determine the SpT, and therefore the effective temperature (T$_{\rm eff}$) of the input spectrum, different values of the extinction and a reddening law to obtain $A_V$, and a set of models which describe the accretion spectrum we use to derive $L_{\rm acc}$. 

{\it Photospheric templates.} The set of photospheric templates is obtained from the one collected in \citet{Manara13}. This is a sample of 24 well-charachterized X-Shooter spectra of non-accreting (Class~III) YSOs representative of objects of SpT classes from late K to M. We extend this grid with two new X-Shooter observations of non-accreting YSOs from the ESO programs 089.C-0840 and 090.C-0050 (PI Manara), one object with SpT G4 and the other one with SpT K2. In total, our photospheric templates grid consists of 26 non-accreting YSOs with SpT between G4 and M9.5. We use Class~III YSOs as photospheric template, because synthetic spectra or field dwarfs spectra would be inaccurate for this analysis for the following reasons. Firstly, YSOs are highly active, and their photosphere is strongly modified by this chromospheric activity, both in the continuum and in the line emission \citep[e.g.][]{Ingleby11,Manara13}. Secondly, field dwarf stars have a different surface gravity with respect to PMS stars, which are sub-giants. Using spectra of non-accreting YSOs as templates mitigates these problems. 

{\it Extinction.} We consider  in the analysis values of $A_V$ in the range [0-10] mag, with steps of 0.1 mag in the range $A_V$ = [0-3] mag, and of 0.5 mag at higher values of $A_V$. This includes all the possible typical values of $A_V$ for non-embedded objects in this region \citep[e.g.][]{DaRio10,DaRio12}. The reddening law we adopt in this work is the one from \citet{Cardelli} with $R_V$=3.1, appropriate for the ONC region \citep{DaRio10}. 

{\it Accretion spectrum.} To determine the excess emission due to accretion, and thus the $L_{\rm acc}$, we use a grid of isothermal hydrogen slab models, which has already been used and proved to be adequate for this analysis \citep[e.g.][]{Valenti93, Herczeg08, Rigliaco12, Alcala13}. We describe the emission due to accretion with the slab model in order to have a good description of the shape of this excess and to correct for the emission arising in the spectral region at wavelengths shorter than the minimum wavelengths covered by the X-Shooter spectra, i.e. $\lambda\lesssim$ 330 nm.
In these models we assume local thermodynamic equilibrium (LTE) conditions, and we include both the H and H$^-$ emission. Each model is described by three parameters: the electron temperature (T$_{\rm slab}$) , the electron density ($n_e$), and the optical depth at $\lambda$=300 nm ($\tau$), which is related to the slab length. The \lacc \ is given by the total luminosity emitted by the slab. The grid of slab models we adopt covers the typical values for the three parameters: T$_{\rm slab}$ are selected in the range from 5000 to 11000 K, $n_e$ varies from 10$^{11}$ to 10$^{16}$ cm$^{-3}$, and $\tau$ has values between 0.01 and 5.

{\it Additional parameters.} In addition to the 5 parameters just introduced, namely the photospheric template, $A_V$, and the three slab model parameters (T$_{\rm slab}$, $n_e$, $\tau$), there is the need to include also two normalization constant parameters, one for the photospheric template ($K_{\rm phot}$) and one for the slab model ($K_{\rm slab}$). The first rescales the emitted flux of the photospheric template to the correct distance and radius of the input target, while the latter converts the slab emission flux as it would have been emitted at the stellar surface by  a region with the area given by the slab parameters. 

\subsection{Multi-component fit: best fit determination}

To consider the three components (SpT, $A_V$, \lacc) altogether, we develop a Python procedure which determines the model which best fits the observed spectrum.  
We calculate for each point of the model grid a likelihood function, which can be compared to a $\chi^2$ distribution, comparing the observed and model spectra in a number of spectral features, which are chosen in order to consider both the region around the Balmer jump, where the emitted flux is mostly originated in the accretion shock, and regions at longer wavelengths, where the photospheric emission dominates the observed spectrum. The form of this function, that will be addressed as $\chi^2_{\rm like}$ function, is the following:
\begin{equation}
\chi^2_{\rm like} = \sum_{features} \left[\frac{ f_{\rm obs} - f_{\rm mod} }{\sigma _{\rm obs}} \right]^2 ,
\label{eq::chi2}
\end{equation}
where $f$ is the value of the measured feature, $obs$ refers to measurements operated on the observed spectrum, $mod$ on those on the model spectrum, and $\sigma_{\rm obs}$ is the error on the value of the feature in the observed spectrum.
The features considered here are the Balmer jump ratio, defined as the ratio between the flux at $\sim$360 nm and that at $\sim$ 400 nm, the slope of the Balmer continuum between $\sim$335 nm and $\sim$ 360 nm, the slope of the Paschen continuum between $\sim$ 400 nm and $\sim$ 475 nm, the value of the Balmer continuum at $\sim$360 nm, and that of the Paschen continuum at $\sim$ 460 nm, and the value of the continuum in different bands at $\sim$ 710 nm. The exact wavelength ranges of these features are reported in Table~\ref{tab::features}. The best fitting model is determined by minimization of the $\chi^2_{\rm like}$ distribution. The exact value of the best fit $\chi^2_{\rm like}$ is not reported, because this value itself should not be considered as an accurate estimate of the goodness of the fit. Whereas a fit with a value of $\chi^2_{\rm like}$ much larger than the minimum one leads to a very poor fit of the observed spectrum, this function is not a proper $\chi^2$, given that it considers only the errors on the observed spectrum and only some regions in the spectrum.

\begin{table}
\centering
\caption{\label{tab::features} Spectral features used to calculate the best-fit in our multi-component fit procedure }
\begin{tabular}{lc}
\hline\hline
Name & Wavelength range   \\
\hbox{} & [nm] \\
\hline
Balmer Jump ratio & (355-360)/(398-402)  \\
Balmer continuum (slope) & 332.5-360 \\
Paschen continuum (slope) & 398-477 \\
Continuum at $\sim$ 360  & 352-358 \\
Continuum at $\sim$ 460 & 459.5-462.5 \\
Continuum at $\sim$ 703 & 702-704 \\
Continuum at $\sim$ 707& 706-708 \\
Continuum at $\sim$ 710 & 709-711 \\
Continuum at $\sim$ 715  & 714-716 \\
\hline
\end{tabular}
\end{table}

The procedure is the following. For each photospheric template we deredden the observed spectrum with increasing values of $A_V$. Then, for every value of $A_V$, considering each slab model, we determine the value of the two normalization constants $K_{\rm phot}$ and $K_{\rm slab}$ by finding the values of these two constants which lead the flux at $\lambda\sim$ 360 nm and at $\lambda\sim$ 710 nm of the normalized sum of the photosphere and the slab model to better match the observed spectrum. 
After that we calculate the $\chi^2_{\rm like}$ value using Eq.~(\ref{eq::chi2}). 
This is done for each point of the grid (SpT, $A_V$, slab parameters).  
After the iteration on each point of the grid terminates, we find the minimum value of the $\chi^2_{\rm like}$ and the correspondent values of the best fit parameters (SpT, $A_V$, slab parameters, $K_{\rm phot}$, $K_{\rm slab}$). 

We also derive from the $\Delta\chi^2_{\rm like}$ distribution with respect to the SpT of the photospheric templates and the different values of $A_V$ the uncertainties on these two parameters, and thus on \lacc. Indeed, these are the main sources of uncertainty in the determination of \lacc, which is in fact a measurement of the excess emission with respect to the photospheric one due to accretion. Most of the accretion excess ($\gtrsim$70\%) is emitted in regions covered by our X-Shooter spectra, and originates mostly in the wavelength range $\lambda\sim 330$-1000 nm, while most of the excess emission at longer wavelengths is due to disk emission and is not considered in our analysis. To derive the total excess due to accretion, we need a bolometric correction for the emission at wavelengths shorter than those in the X-Shooter range, i.e. $\lambda\lesssim$330 nm. This is calculated with the best fit slab model. Analyzing the slab models we derive that this contribution accounts for less than 30\% of the excess emission, and that the shape of this emission is well constrained by the Balmer continuum slope. Different slab model parameters with reasonable Balmer continuum slopes, that would imply similarly good fits as the best one, would lead to values of \lacc \ always within 10\% of each other, as it has been pointed out also by \citet{Rigliaco12}. Therefore, once SpT and $A_V$ are constrained, the results with different slab parameters will be similar. With our procedure we can constrain very well all the parameters. Typically, the solutions with $\chi^2_{\rm like}$ values closer to the best-fit one are those within one spectral subclass of difference in the photospheric template, $\pm$0.1-0.2 mag in extinction values, and with differences of \lacc \ of less than $\sim$10\%. Other sources of uncertainty on the estimate of \lacc \ are the noise in the observed and template spectra, the uncertainties in the distances of the target, and the uncertainty given by the exclusion of emission lines in the estimate of the excess emission (see e.g. \citealt{Herczeg08,Rigliaco12,Alcala13}). Altogether, typical errors on estimates of \lacc \ with our method are of 0.2-0.3 dex.

As previously mentioned, the accretion emission veils the photospheric absorption features of the observed YSO spectrum. Those can be used to visually verify the goodness of the best fit obtained with the $\chi^2_{\rm like}$ minimization procedure. In order to do that, we always plot the observed spectrum and the best fit one in the Balmer jump region, in the Ca~I absorption line region at $\lambda\sim$420 nm, in the continuum bands at $\lambda \sim$ 710 nm, and in the spectral range of different photospheric absorption lines which depends mainly on the temperature of the star (T$_{\rm eff}$), namely TiO lines at $\lambda\lambda$ 843.2, 844.2, and 845.2 nm, and the Ca~I line at $\lambda\sim$ 616.5 nm. The agreement between the best fit and the observed spectrum in the Balmer jump region, at $\lambda\sim 420$ nm, and in the continuum bands at $\lambda\sim$ 710 nm is usually excellent. Similarly, the best fit spectrum reproduces very well the observed one also in the photospheric features at longer wavelengths. 

\subsection{Comparison to synthetic spectra}
As an additional check of our results, and to derive the surface gravity (log$g$) for the target, we compare its spectrum with a grid of synthetic spectra. In particular, we adopt synthetic BT-Settl spectra \citep{Allard11} with solar metallicity, log$g$ in the range from 3.0 to 5.0 (in steps of 0.5), and T$_{\rm eff}$ equal to that of the best fit photospheric template, using the SpT-T$_{\rm eff}$ relation from \citet{Luhman03}, and also with T$_{\rm eff}$ correspondent to the next upper and lower spectral sub-classes. 

The procedure we use is the following. First, we correct the input spectrum for reddening using the best fit value of $A_V$. Subsequently, we remove the effect of veiling by subtracting the best fit slab model, rescaled with the constant $K_{\rm slab}$ derived in the fit, to the dereddened input spectrum. Then, we degrade the synthetic spectra to the same resolution of the observed one (R=17400 in the VIS), and we resample those to the same wavelength scale of the target. We then select a region, following \citet{Stelzer13}, with many strong absorption lines dependent only on the star T$_{\rm eff}$, and with small contamination from molecular bands and broad lines. This is chosen to derive the projected rotational velocity ($v$sin$i$) of the target by comparison of its spectrum with rotationally broadened synthetic spectra. The region chosen is in the VIS arm, from 960 to 980 nm, and is charachterized by several Ti~I absorption lines, and by some shallower Cr~I lines. Then, we broaden the synthetic spectra to match the $v$sin$i$ of the observed one, and we compare it with the reddened- and veiling-corrected target spectrum in two regions, which are chosen, always following \citet[][and references therein]{Stelzer13}, because they include absorption lines sensitive to both $T_{\rm eff}$ and log$g$ of the star. These are one in the wavelength range from 765 to 773 nm, where the K~I doublet ($\lambda\lambda\sim$ 766-770 nm) is, and one in the range from 817 to 822 nm, where the Na~I doublet ($\lambda\lambda\sim$ 818.3-819.5 nm) is. The best fit log$g$ is the one of the synthetic spectrum with smaller residuals oberved-synthetic in the line regions. 

\subsection{Stellar parameters}

From the best fit parameters determined with the procedure explained in this Section we obtain T$_{\rm eff}$, $A_V$, and \lacc. The first corresponds to the T$_{\rm eff}$ of the best fit photospheric template, converted from the SpT using the \citet{Luhman03} relation, derived with the multi-component fit and verified with the comparison with the synthetic spectra. $A_V$ is also derived in the multi-component fit,  while \lacc \ is calculated by integrating from 50 nm to 2500 nm the total flux of the best fit slab model, rescaled using the normalization constant $K_{\rm slab}$ derived before. This total accretion flux ($F_{\rm acc}$) is then converted in \lacc \ using the relation $L_{\rm acc} = 4\pi d^2 F_{\rm acc}$, where $d$ is the distance of the target. 

To derive $L_*$ of the input spectrum we use the values of $L_*$ of the non-accreting YSOs used as photospheric template for our analysis, which have been derived in \citet{Manara13}. From the best fit result, we know that $f_{\rm obs, dereddened} = K_{\rm phot}\cdot f_{\rm phot} + K_{\rm slab}\cdot f_{\rm slab}$, where $f$ is the flux of the spectrum of the dereddened observed YSO ($obs,dereddened$), of the photospheric template ($phot$), and of the slab model ($slab$). The photospheric emission of the input target is then given solely by the emission of the photospheric template rescaled with the constant $K_{\rm phot}$. Therefore, we use the relation: 
\begin{equation}
L_{*,\rm obs} = K_{\rm phot} \cdot (d_{\rm obs}/d_{\rm phot})^2 L_{*,\rm phot},
\end{equation}
where $d$ is the distance, $L_{*,\rm phot}$ is the bolometric luminosity of the photospheric template, and $L_{*,\rm obs}$ that of the input object. With this relation we derive $L_{*}$ for the input YSO. 

From T$_{\rm eff}$ and $L_*$ we then derive $R_*$, whose error is derived by propagation of the uncertainties on T$_{\rm eff}$ and $L_*$. $M_*$ and age are obtained by interpolation of evolutionary models \citep{Siess00,Baraffe98,Palla,DAntona} in the position of the target on the HRD. Their typical uncertainties are determined allowing their position on the HRD varying according to the errors on T$_{\rm eff}$ and $L_*$. Finally, using the relation $\dot{M}_{\rm acc} = 1.25\cdot L_{\rm acc}R_*/(GM_*)$ \citep{Gullbring98} we derive \macc, and its error is determined propagating the uncertainties on the various quantities in the relation. We report always the values derived using all the evolutionary models, in order to show that the results are not model-dependent.

Another important quantity which can be used to asses the young age of a PMS star is the presence and the depth of the lithium absorption line at $\lambda\sim$ 670.8. Given that veiling modifies substantially the equivalent width of this line (EW$_{\rm Li}$), we need to use the reddening- and veiling-corrected spectrum of the target to derive this quantity. On this corrected spectrum we calculate EW$_{\rm Li}$ integrating the gaussian fit of the line, which is previously normalized to the local pseduo-continuum at the edges of the line. The statistical error derived on this quantity is given by the propagation of the uncertainty on the continuum estimate. 

\section{Results}
\label{sect::results}

In the following we report the accretion and stellar properties of the two older PMS candidates obtained using the procedure explained in Sect.~\ref{sect::method}. 


\begin{figure*}
\centering
\includegraphics[width=0.7\textwidth]{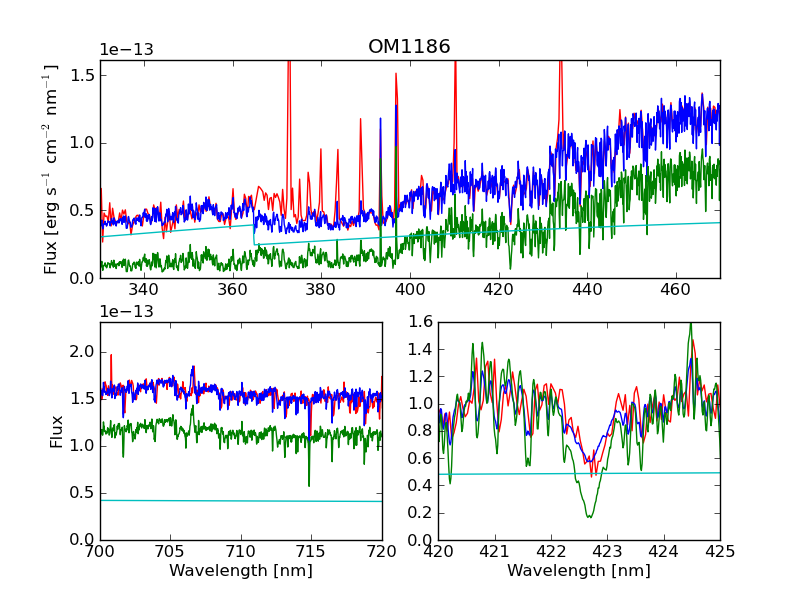}
\includegraphics[width=0.6\textwidth]{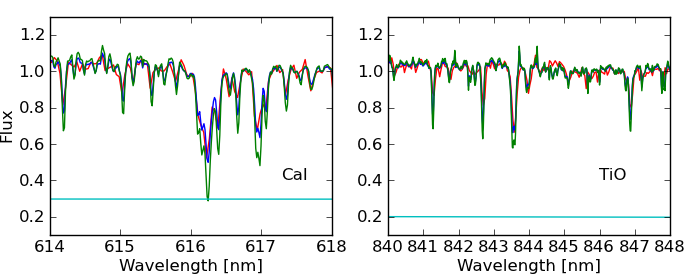}
\caption{Best fit for the object OM1186. Balmer jump, normalization region, CaI absorption feature, and photospheric features used to check derived veiling. }
        \label{fig::OM1186_acc}
\end{figure*}

\subsection{OM1186}
The best fit for OM1186 is shown in Fig.\ref{fig::OM1186_acc}: the red line is the observed spectrum, the green one the photospheric template, the cyan line the slab model, and the blue the best fit. The region around the Balmer jump is shown in the upper plot, while the other panels show the normalization region around $\sim$710 nm, the continuum normalized Ca~I absorption line at $\sim$ 422 nm, and the photospheric features at $\lambda\sim$ 616 nm and $\lambda\sim$ 844 nm. We clearly see that the agreement between the observed and best fit spectra is very good in all the region analyzed. This is obtained using a photospheric template of SpT K5, which corresponds to T$_{\rm eff}$=4350 K with an estimated uncertainty of 350 K, $A_V$ = 0.9$\pm$0.4 mag, and slab parameters leading to a value of \lacc=0.20$\pm$0.09 \lsun. We then confirm the SpT reported in the literature for this object, but we derive a different value for the extinction, which was found to be $A_V$=2.4 mag.  

\begin{figure*}
\centering
\includegraphics[width=0.7\textwidth]{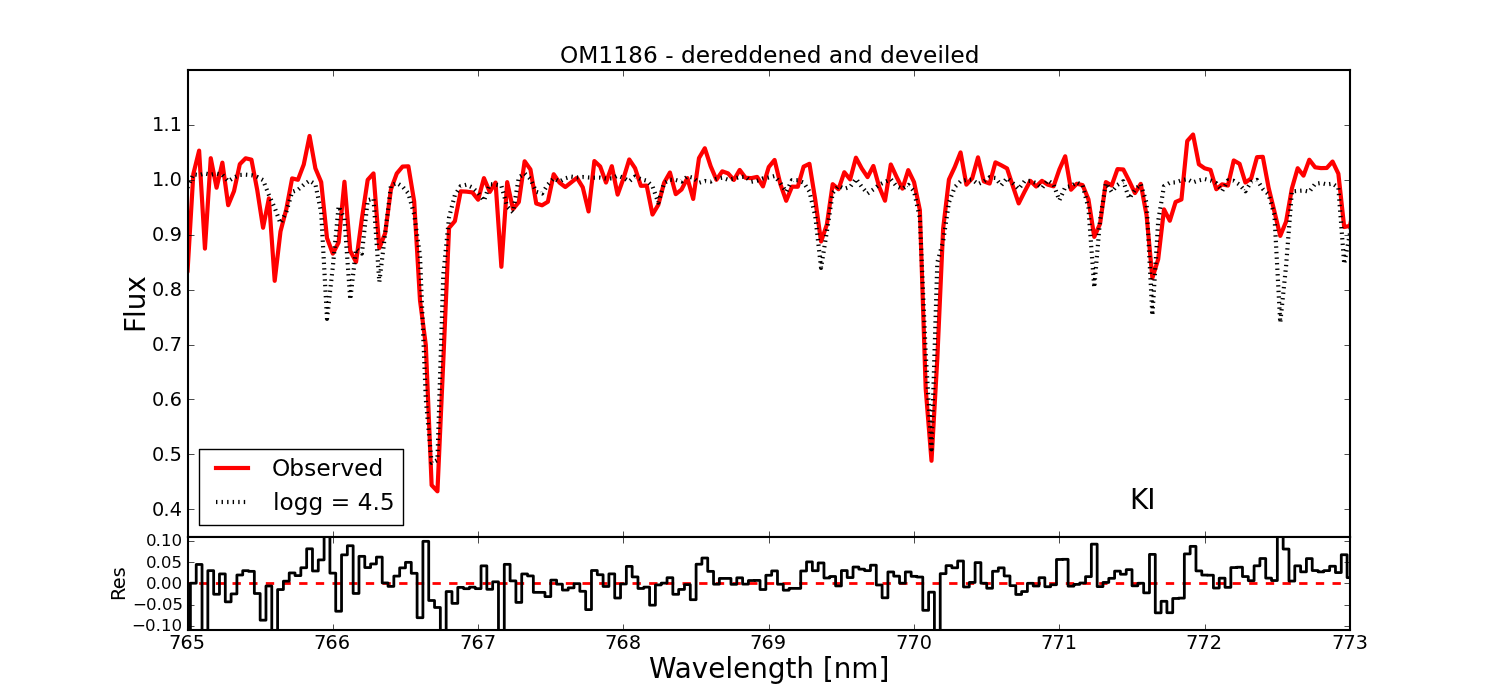}
\includegraphics[width=0.7\textwidth]{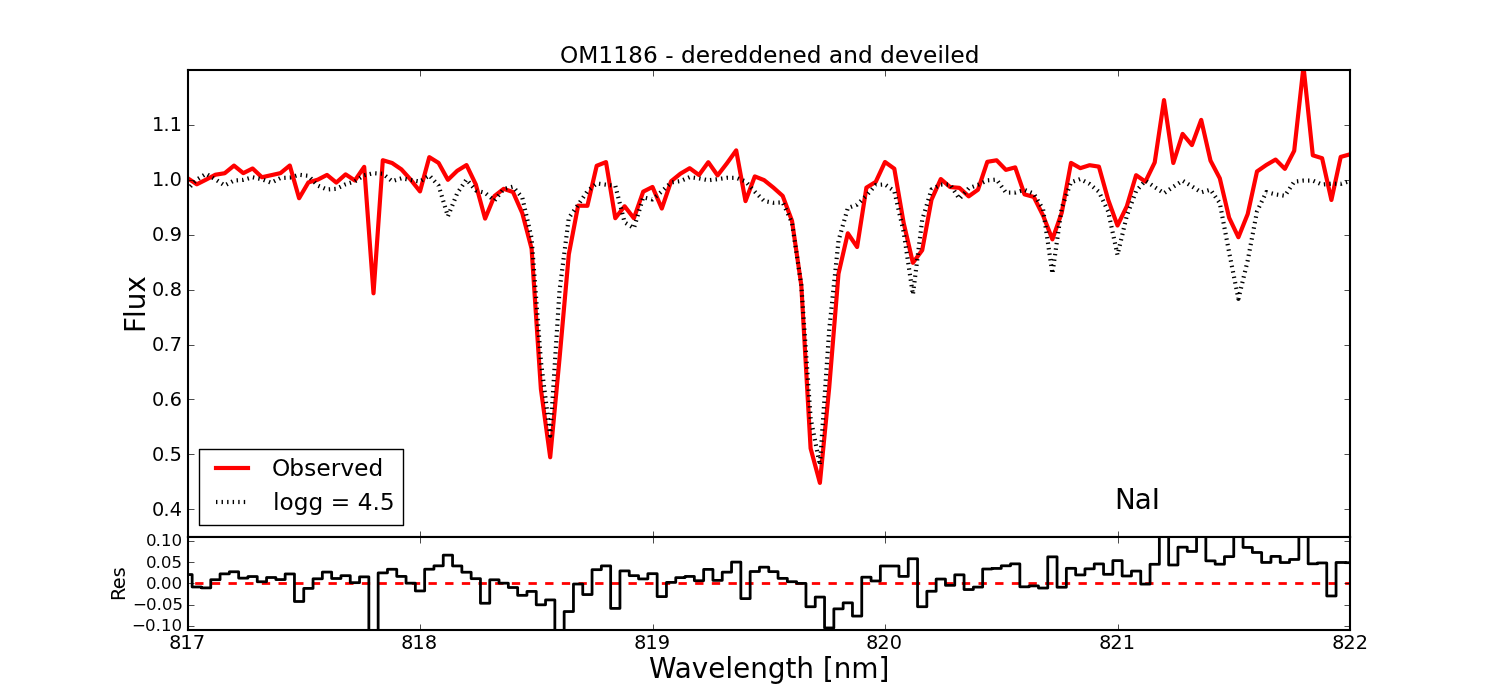}
\caption{Comparison of the reddening- and veiling-corrected spectrum of OM1186 with a synthetic spectrum with T$_{\rm eff}$ = 4350 K and log$g$=4.5. }
        \label{fig::OM1186_synt}
\end{figure*}

The result of the comparison of the input spectrum with the grid of synthetic spectra is shown in Fig.~\ref{fig::OM1186_synt}. The best agreement is found using a synthetic spectrum at the same nominal T$_{\rm eff}$ of the best fit photosphere, i.e. T$_{\rm eff}$ = 4350 K, and with log$g$=4.5. In Fig.~\ref{fig::OM1186_synt} both regions adopted for the log$g$ analysis are shown, and the red line refers to the reddening- and veiling-corrected observed spectrum, while the black dotted line is the synthetic spectrum which better reproduces the observed one. We also report the residuals in the bottom panel.


From the best fit, we derive for this object a value of $L_*$ = 1.15$\pm$0.36 \lsun. With this value and using T$_{\rm eff}$ = 4350 K, we derive the mass and age for these objects with the evolutionary models of \citet{Siess00,Baraffe98,Palla,DAntona}. The values derived are, respectively, $M_*$= \{1.1$\pm$0.4, 1.4$\pm$0.3, 1.1$\pm$0.4, 0.6$\pm$0.3\} \msun \ and age = \{3.2$^{+4.8}_{-2.0}$, 4.7$^{+7.2}_{-2.5}$, 2.8$^{+2.9}_{-1.7}$, 0.8$^{+1.4}_{-0.4}$\} Myr. Moreover, we derive \macc\ with the parameters derived from the fit and from the evolutionary models and, according to the different evolutionary tracks, we obtain \macc=\{1.3 $\pm$0.8, 1.1$\pm$0.6, 1.4$\pm$0.8, 2.4$\pm$1.8\} $\cdot 10^{-8} M_\odot$/yr for this object. From the reddening- and veiling-corrected spectrum we derive EW$_{\rm Li}$ = 658 $\pm$ 85 m\AA. 
All the parameters derived from the best fit are reported in Table~\ref{tab::our}, while those derived using the evolutionary models in Table~\ref{tab::ev_mod}.

\subsection{OM3125}
\begin{figure*}
\centering
\includegraphics[width=0.7\textwidth]{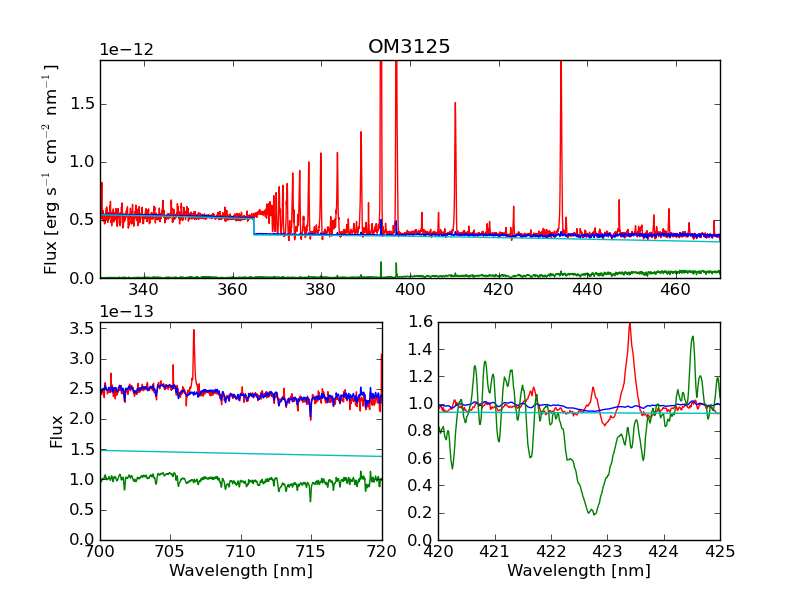}
\includegraphics[width=0.62\textwidth]{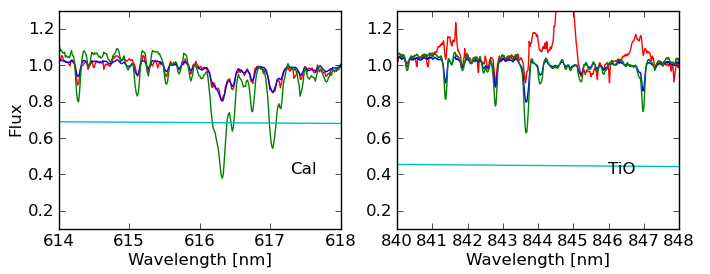}
\caption{Best fit for the object OM3125. Same as Fig.~\ref{fig::OM1186_acc}.}
        \label{fig::OM3125_acc}
\end{figure*}

For OM3125 the values reported in the literature are SpT G8-K0 and $A_V$=1.47 mag. Using the same procedure, we obtain for OM3125 a minimum $\chi^2_{\rm like}$ value with a photospheric template of SpT K7, corresponding to T$_{\rm eff}$=4060 K, with an uncertainty of 250 K, and $A_V$ = 1.2$\pm$0.3 mag. However, with this best fit model we do not reproduce very well the Ca~I absorption feature at $\lambda\sim$616.5 nm, because the amount of veiling in this line is too high. Looking at the solutions with values of $\chi^2_{\rm like}$ similar to the minimum value, we find that the best agreement between the observed and fitting spectrum in this feature is with a value of $A_V$ = 1.0 mag. Please note that the choice of this solution instead of the one with $A_V$ = 1.2 mag implies, well within the errors, the same derived parameters $L_*$ and \lacc. We show this adopted best fit in Fig.~\ref{fig::OM3125_acc}, using the same color-code as in Fig.~\ref{fig::OM1186_acc}. The slab model used here leads to a value \lacc = 1.25$\pm$0.60 \lsun. The effect of veiling in this object is stronger than in OM1186, and this is clearly seen both in the Balmer jump excess and in the Ca~I absorption feature at $\lambda\sim$420 nm, which is almost completely veiled, and has emission on reversal of the very faint absorption feature. Moreover, also the other photospheric features are much more veiled, as it is shown in the bottom panels of Fig.~\ref{fig::OM3125_acc}. In the bottom right panel there are also hydrogen and helium emission lines due to accretion in correspondence with the TiO absorption features normally present in the photosphere of objects with this SpT (see Fig.~\ref{fig::OM1186_acc} for comparison). 


\begin{figure*}
\centering
\includegraphics[width=0.7\textwidth]{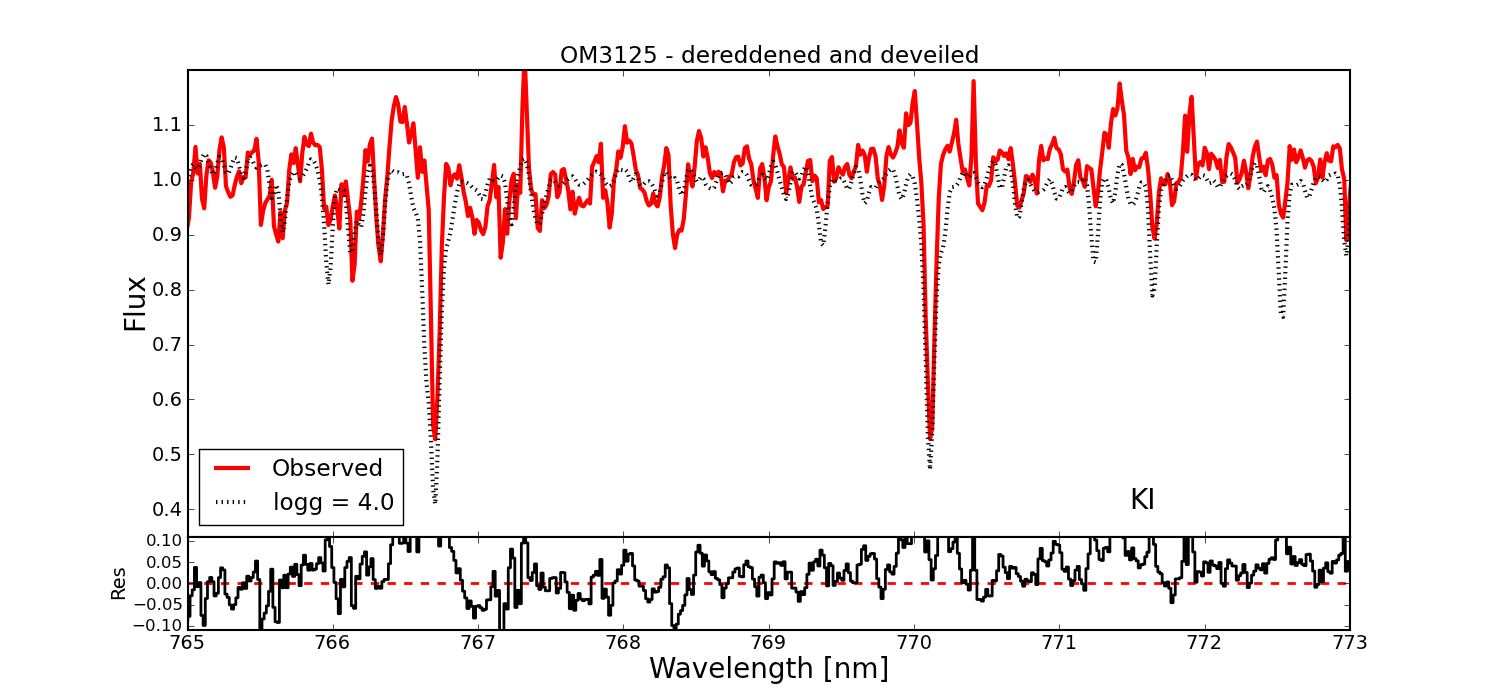}
\includegraphics[width=0.7\textwidth]{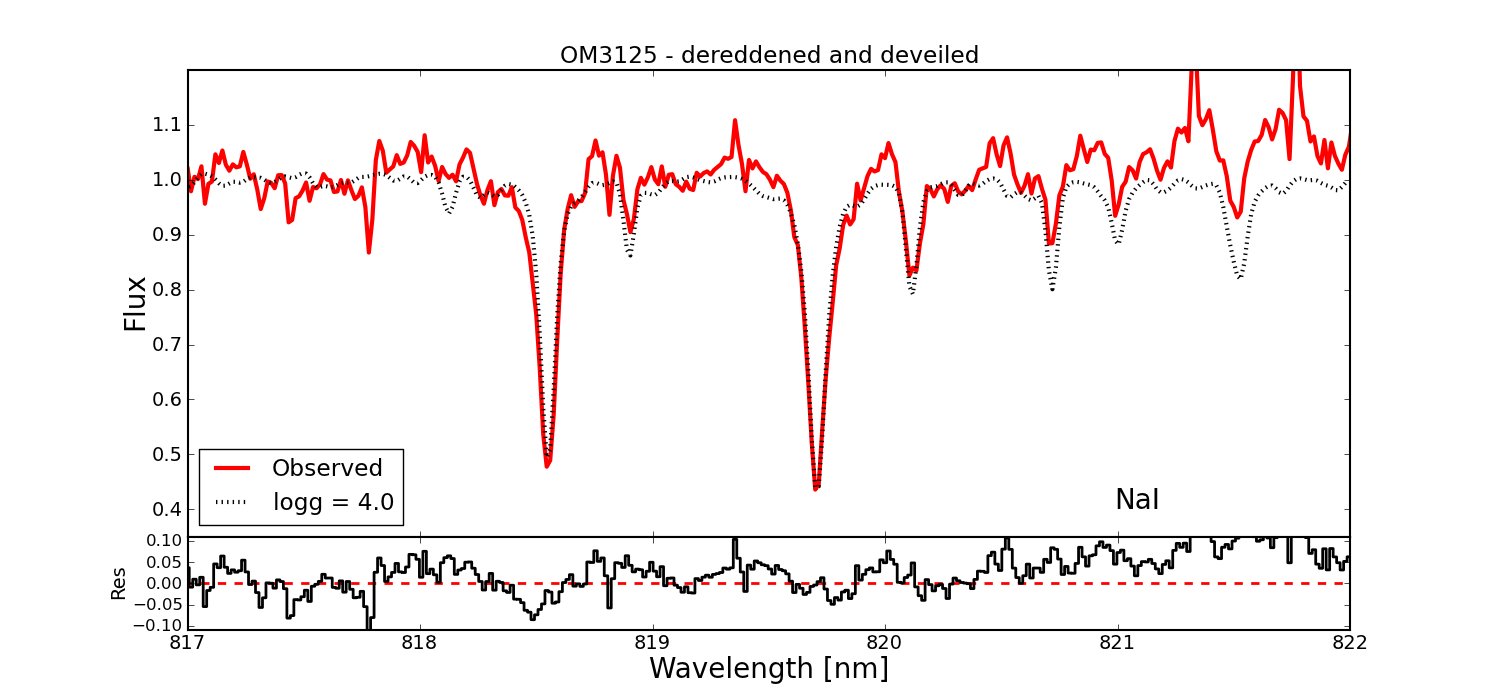}
\caption{Comparison of the reddening- and veiling-corrected spectrum of OM3125 with a synthetic spectrum with T$_{\rm eff}$ = 4000 K and log$g$=4.0. Same as Fig.~\ref{fig::OM1186_synt}.}
        \label{fig::OM3125_synt}
\end{figure*}

In Fig.~\ref{fig::OM3125_synt} we show, using the same color code as in Fig.~\ref{fig::OM1186_synt}, the synthetic spectrum analysis for this object. In this case the best agreement is found with a synthetic spectrum with T$_{\rm eff}$ = 4000 K and with log$g$=4.0. 

%

The luminosity derived for this object from its best fit is $L_*$ = 0.81$\pm$0.44 \lsun. With this value and using T$_{\rm eff}$ = 4060 K, we derive the mass and age for these objects with the evolutionary models of \citet{Siess00,Baraffe98,Palla,DAntona}. The values derived are, respectively, $M_*$= \{0.8$\pm$0.3, 1.2$\pm$0.2, 0.8$\pm$0.2, 0.5$\pm$0.2\} \msun \ and age = \{2.2$^{+6.6}_{-1.0}$, 4.32$^{+8.7}_{-2.0}$, 2.4$^{+4.3}_{-1.3}$, 0.8$^{+2.7}_{-0.3}$\} Myr.  This object has, according to the different evolutionary models, \macc=\{1.2$\pm$0.9, 0.8$\pm$0.5, 1.2$\pm$0.7, 1.9$\pm$1.2\} $\cdot 10^{-7} M_\odot$/yr, and EW$_{\rm Li}$ = 735 $\pm$ 42 m\AA. 
These results are also reported in  Table~\ref{tab::our} and ~\ref{tab::ev_mod}.

\begin{table*}
\centering
\caption{\label{tab::our}Newly derived parameters from the multi-component fit}
\begin{tabular}{lcccccccc}
\hline\hline
Name & SpT   & T$_{\rm eff}$&A$_V$ & L$_*$ & R$_*$ & log$g$ & EW$_{\rm Li}$&\lacc  \\
\hbox{} & \hbox{}  &  [K]  & [mag] & [L$_\odot$] & [R$_\odot$] & [cm/s$^2$] & [m\AA] &[L$_\odot$] \\
\hline
OM1186 & K5 & 4350$\pm$350 & 0.9$\pm$0.4 & 1.15$\pm$0.36 & 1.9$\pm$0.4 & 4.5$\pm$0.5 & 658$\pm$85 & 0.20$\pm$0.09 \\
OM3125 & K7 & 4060$\pm$250 & 1.0$\pm$0.3 & 0.81$\pm$0.44 & 1.8$\pm$0.5 & 4.0$\pm$0.5 & 735$\pm$42 & 1.25$\pm$0.60 \\
\hline
\end{tabular}
\end{table*}

\begin{table*}
\centering
\caption{\label{tab::ev_mod}Parameters derived from evolutionary models using newly derived photospheric parameters}
\begin{tabular}{ l | ccc | ccc }
\hline\hline
 & \multicolumn{3}{c}{OM1186} & \multicolumn{3}{c}{OM3125}\\
Evolutionary & M$_*$& age &\macc & M$_*$& age &\macc \\
model & [M$_\odot$] &[Myr] & [10$^{-8}  M_\odot$/yr] & [M$_\odot$] &[Myr] & [10$^{-8}  M_\odot$/yr] \\
\hline
\citet{Siess00} & 1.1$\pm$0.4 & 3.2$^{+4.8}_{-2.0}$ & 1.3$\pm$0.8 &  0.8$\pm$0.3 & 2.2$^{+6.6}_{-1.0}$ & 12.0$\pm$8.6 \\
\citet{Baraffe98} & 1.4$\pm$0.3 & 4.7$^{+7.2}_{-2.5}$ & 1.1$\pm$0.6 & 1.2$\pm$0.2 & 4.3$^{+8.7}_{-2.0}$ & 7.9$\pm$4.7\\
\citet{Palla} & 1.1$\pm$0.4 & 2.8$^{+2.9}_{-1.7}$ & 1.4$\pm$0.8 & 0.8$\pm$0.2 & 2.4$^{+4.3}_{-1.3}$ & 11.6$\pm$7.5 \\
\citet{DAntona} & 0.6$\pm$0.3 & 0.8$^{+1.4}_{-0.4}$ & 2.4$\pm$1.8 & 0.5$\pm$0.2 & 0.8$^{+2.7}_{-0.3}$ & 18.6$\pm$12.3\\
\hline
\end{tabular}
\end{table*}

\section{Discussion}
\label{sect::discussion}

\begin{figure}
\centering
\includegraphics[width=0.5\textwidth]{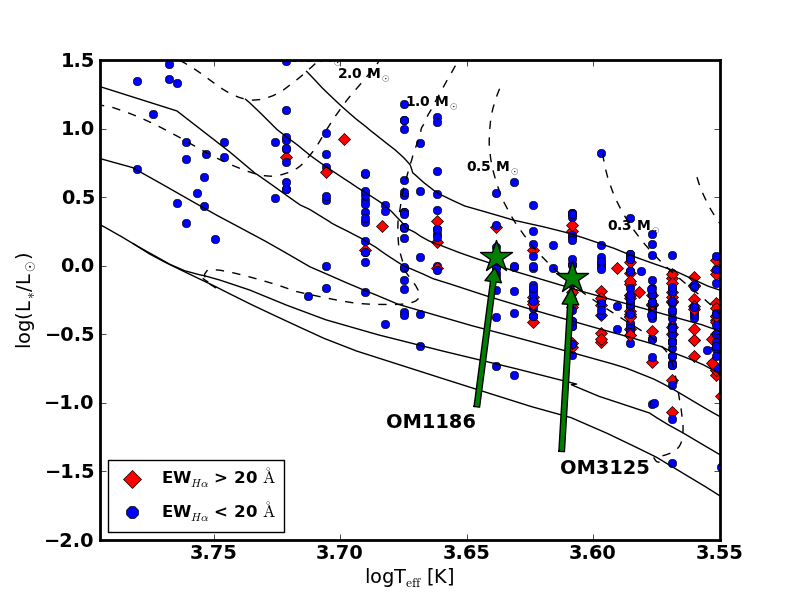}
\caption{H-R Diagram of the ONC from \citet{DaRio12}, with colored stars showing the new positions of the two targets of this study. The overplotted evolutionary tracks are from \citet{DAntona}. We plot (from top to bottom) the 0.3, 1, 3, 10, 30, and 100 Myr isochrones.  }
        \label{fig::HRD_after}
\end{figure}

With the results presented in the previous section, we determine new positions for our target on the HRD. This is shown in Fig.~\ref{fig::HRD_after} with green stars representing the two YSOs analyzed in this study and the other symbols the rest of the ONC population, as in Fig.~\ref{fig::HRD}. Their revised positions are compatible with the bulk of the population of the ONC, and their revised mean ages, i.e. $\sim$2.9 Myr for OM1186 and $\sim$2.4 Myr for OM3125, are typical ages for objects in this region, which mean age has been estimated around 2.2 Myr \citep{Reggiani11}. We check also that the final results do not depend on the value we have chosen for the reddening law, i.e. $R_V$=3.1. With a value of $R_V$=5.0 we obtain ages which are systematically younger than the one obtained in our analysis by a factor $\sim$30\% for OM1186 and $\sim$50\% for OM3125. With the newly determined parameters, these objects are clearly not older than the rest of the population, and their status of candidate older PMS is due to an incorrect estimate of the photospheric parameters in the literature. Fig.~\ref{fig::HRD_after} also shows that most of the objects which then appear older than 10 Myr have small or negligible H$\alpha$ excess (blue circles). Among the objects with age $\gtrsim$ 10 Myr, only 8 objects actually show signatures of intense accretion (one visible in the plot as a red diamond, the other 7 have T$_{\rm eff} < 3550$ K and are, therefore, outside the plot range) and, at the same time, of age older than 10 Myr. These objects should be observed in the future with techniques similar to the one we used here (see Sect.~\ref{subsec::implications} for details) to understand their real nature.  
In the following, we will analyze other derived parameters for these objects which confirm the age estimated with the HRD. Then, we will address possible reasons which lead to misclassification of these targets in the literature. 

\subsection{Age related parameters}

{\it Lithium abundance:} using the values of EW$_{\rm Li}$ and the stellar parameters obtained in the previous section we calculate the lithium abundance (log$N$(Li)) for the two targets by interpolation of the curves of growth provided by \citet{Pavlenko96}. We derive log$N$(Li) = 3.324$\pm$0.187 for OM1186 and log$N$(Li)=3.196$\pm$0.068 for OM3125. These values are compatible with the young ages of the targets according to various evolutionary models \citep{Siess00,DAntona,Baraffe98}. Indeed, these models predict almost no depletion of Lithium for objects with these T$_{\rm eff}$ at ages smaller than 3 Myr, meaning that the measured lithium abundance for younger objects should be compatible with the interstellar abundance log$N$(Li)$\sim$3.1-3.3 \citep{Palla07}. 

 {\it Surface gravity: } The derived values of the surface gravity for the two targets are compatible with the theoretical values for objects with T$_{\rm eff}\sim$4000-4350 K and an age between $\sim$1-4.5 Myr. Indeed, both \citet{Siess00} and \citet{Baraffe98} models predict a value of log$g$ $\sim$ 4.0 for objects with these properties. Nevertheless, the derived value of log$g$ for OM1186 is also typical of much older objects, given that models predict log$g$ $\sim$ 4.5 at ages $\sim$20 Myr, with very small increase in the next evolutionary stages. However, the uncertainties on the determination of this parameter are not small, and the increase of this value during the PMS phase is usually within the errors of the measurements.

\subsection{Sources of error in the previous classifications}
Both targets have been misplaced on the HRD, but the reasons were different. For OM1186 we have confirmed the SpT available from the literature, but we found different values of $A_V$ and \lacc \ with respect to \citet{DaRio10}. In their work, they use a color-color $BVI$ diagram to derive simultaneously $A_V$ and \lacc, with the assumption of the SpT. To model the excess emission due to accretion, they use a superposition of an optically thick emission, which reproduces the heated photosphere, and of an optically thin emission, which models the infalling accretion flow. From this model spectrum they derive the contribution of accretion to the photometric colors of the targets by mean of synthetic photometry. Their method assumes that on the $BVI$ diagram the positions of the objects are displaced from the theoretical isochrone due to a combination of extinction and accretion. With the assumption of the SpT, they can find the combination of parameters ($A_V$, \lacc) which best reproduces the positions of each target on the $BVI$ diagram. With these values, they corrected the $I$-band photometry for the excess due to accretion, derived using the accretion spectrum model, and for reddening effects. Finally, they derived $L_*$ from this corrected $I$-band photometry using a bolometric correction. Using this method, they found a solution for OM1186 with a large value of $A_V$ and, subsequently, of accretion. Given the large amount of excess due to accretion they derive in the $I$-band, they underestimated $L_*$ and assigned an old age to this target. On the other hand, our revised photospheric parameters are compatible with those of \citet{Manara12}, who found $A_V$=1.16 mag, $L_*$=0.77 $L_\odot$, and age$\sim$6 Myr. In their analysis, they used the same method as \citet{DaRio10}, but they had at disposal also $U$-band photometric data, and thus used an $UBI$ color-color diagram. Given that the excess emission due to accretion with respect to the photosphere in the $U$-band is much stronger, they were able to determine the accretion properties of the targets more accurately, and to find a unique correct solution. 

Regarding OM3125, we derived with our analysis a different SpT with respect to the one usually assumed in the literature \citep[G8-K0;][]{Hillenbrand97}. This estimate was obtained using an optical spectrum covering the wavelength region from $\sim$500 nm to $\sim$900 nm, and their analysis did not consider the contribution of accretion to the observed spectrum. This was assumed not to be a strong contaminant for the photospheric features in this wavelength range. This is a reasonable assumption for objects with low accretion rates, but it has been later shown not to be accurate for stronger accretors \citep{Fischer11}, as we already pointed out for this target. In particular, strong veiling makes the spectral features shallower, which lead to an incorrect earlier classification of the target. \citet{Hillenbrand97} marked this SpT classification as uncertain, and they reported also the previous classification for this object from \citet{CK79}, who classified it as being of SpT K6, a value which is in agreement with our finding. This previous classification is obtained using  spectra at shorter wavelengths ($\lambda\sim$ 420-680 nm) with respect to \citet{Hillenbrand97}, and still with no modeling of the accretion contribution. The difference in the classification is most likely due to the fact that some TiO features at $\lambda\lambda\sim$476, 479 nm, which are typical of objects of spectral class late-K, 
were covered in the spectra of \citet{CK79}. Given the large amount of veiling due to accretion that is present in the spectrum of this object, it represents a clear case where the detailed analysis carried out in our work is needed. We conclude that even optical spectra covering large regions of the spectra, like those used in \citet{Hillenbrand97} and in \citet{CK79}, can lead to different, and sometimes incorrect, results if the veiling contribution is not properly modeled.

\subsection{Implications of our finding}
\label{subsec::implications}

Accretion veiling, extinction and spectral type are difficult to estimate accurately from limited datasets, especially if mostly based on only few photometric bands. In particular, when classifying young stellar objects the difficulty increases, given that accretion and extinction may modify substantially the observed spectra. For this reason, the use of few photometric bands where accretion, extinction and spectral type effects can be very degenerate (e.g. $B$-$I$ band range) or of spectra covering only small wavelength regions (e.g. $\lambda\sim$500-900 nm) can lead to different solutions, which may be incorrect.

With our work we show that an analysis of the whole optical spectrum which includes a detailed modeling of the various components of the observed spectrum leads to an accurate estimate of the stellar parameters. When good quality and intermediate resolution spectra with large wavelength coverage, i.e. from $\lambda\sim$330 nm to $\lambda\sim$ 1000 nm, are not available, we suggest that also a combination of photometric and spectroscopic data could lead to a robust estimate of the stellar parameters of the targets independently of their SpT. In particular, the photometric data should cover the region around the Balmer jump (using both $U$- and $B$-band), and that around $\lambda\sim$700 nm ($R$- and $I$-band). With this dataset it would be possible to derive accretion and veiling properties, extinction, and SpT by means of photometrical methods similar to those of \citet{Manara12}, but expanded to include different photospheric templates, and different accretion spectra. At the same time, spectra in selected wavelength regions where photospheric lines sensitive to T$_{\rm eff}$ and/or log$g$ are present are necessary to pin-down degeneracies on the SpT. Finally, the lithium absorption line should be included as a further test of the age.

Our work also implies that single objects which deviate from the bulk of the population in nearby young clusters (age $\lesssim$ 3 Myr) could be affected by an incorrect estimate of the photospheric parameters especially in cases where the determination is based on few photometric bands and the effects of veiling due to accretion and extinction are strong.  
On the other hand, we have no reason to believe that also a large number of objects positioned along the isochrones representative of the bulk of the population in nearby regions should be affected by similar problems. We think that the vast majority of the estimates available are correct for the following reasons. First of all, objects with low or negligible accretion should be easier to classify given that their spectral features are not affected by veiling. Then, in various regions there are small effects due to differential extinction. In particular, when dealing with objects located in regions less affected by extinction effects, e.g. $\sigma$-Ori, we expect the amount of misclassified objects to be particularly small. 
Larger numbers of objects could be misclassified in very young regions (age$\lesssim$1 Myr, e.g. $\rho$-Oph), where both accretion and extinction effects are strong. Here we could have both underestimation of \lstar, like in the case of OM1186, due to an overestimation of \lacc, and the opposite effect of an overestimation of \lstar \ due to an underestimate of \lacc. These effects could contribute to the spread of \lstar \ which is observed in almost all star forming regions. 
Finally, accretion variability effects should not affect substantially the properties of most of the targets, given that these effects are usually small \citep[e.g.][]{Costigan12}.  

Even though any conclusion on the nature of the older populations observed in massive clusters cannot be drawn from this work, we suggest that the method explained here, or the alternative approach suggested in this section, should be used to study these objects in more detail.  

\section{Conclusion}
\label{sect::conclusion}

We have observed with the ESO/VLT X-Shooter spectrograph two candidate old (age$>$10Myr) accreting PMS in the ONC in order to confirm previous accretion rate and age estimates. Using a detailed analysis of the observed spectra based on a multi-component fit, which includes the photospheric emission, the effect of reddening, and the continuum excess due to accretion, we derived revised accretion rates and stellar parameters for the two targets. We confirm that the objects are accretors as from previous studies, but the revised photospheric parameters place these objects in the same location as the bulk of the ONC young stellar population (age$\sim$2-3Myr). 
Therefore, they cannot be considered older PMS, but they are classical accreting YSOs. In particular, we confirmed the previously estimated SpT for OM1186, but we derived different values of $A_V$ and \lacc \ with respect to previous works in the literature, finding that the real position of this object on the HRD leads to a mean age estimate of $\sim$2.9 Myr. On the other hand, we derived a different SpT for OM3125 with respect to the usually assumed value in the literature. This, together with the other parameters, moved this target to a position on the HRD leading to age $\sim$2.4 Myr. The analysis of the lithium abundance confirms this finding, and that of the surface gravity also, even if the uncertainties for the latter are large.

We showed that with our analysis we are able to accurately determine the stellar parameters of YSOs, while the use of few photometric bands alone (e.g. only the $B$- and $I$-band) or only spectra covering small wavelength regions can lead to large errors in the derived position on the HRD. We thus suggest that in nearby young clusters single objects whose position on the HRD is not compatible with the bulk of the population in their region could be misplaced, especially if there is the suspicion of high optical veiling connected to large values of the accretion rate. 
The nature of these individual objects will need to be verified using a detailed analysis similar to the one we report in this study, in order to verify the existence and study the properties of long-lived disks around young stellar objects.

\begin{acknowledgements}
We thank the anonymous referee for providing useful comments which helped us to improve the paper. We thank the ESO Director General for awarding DDT time to this project and the ESO staff in Paranal for carrying out the observations in Service mode. We thank Andrea Banzatti for useful comments and discussion which helped to improve the fitting procedure. C.F.M. acknowledges the PhD fellowship of the International Max-Planck-Research School. G.B. acknowledges the European Community's Seventh Framework Programme under grant agreement no. 229517.
\end{acknowledgements}

\end{document}